\pdfoutput=1

\documentclass[letterpaper, 10 pt, conference]{ieeeconf}  

\IEEEoverridecommandlockouts                              

\usepackage{graphicx}        

\usepackage{algorithm} 
\usepackage{algorithmic}
\usepackage{multirow}
\usepackage{amsmath}
\usepackage{xcolor}
\usepackage{subfigure}

\title{\LARGE \bf
Spatial and Spectral Features Fusion for EEG Classification during Motor Imagery in BCI
}

\author{Chuanqi Tan$^{1,2}$, Fuchun Sun$^{1,*}$, Wenchang Zhang$^{1}$, Shaobo Liu$^{1}$, Chunfang Liu$^{1}$
\\$^{1}$Department of Computer Science and Technology, Tsinghua University, Beijing, China
\thanks{$^{2}$Chuanqi Tan is a Ph.D candidate in Department of Computer Sciences and Technology, Tsinghua University,
        100084, Beijing, China
        {\tt\small tcq15@mails.tsinghua.edu.cn}}
\thanks{$^{*}$Corresponding author: \tt\small fcsun@mail.tsinghua.edu.cn}
}

\begin{document}

\maketitle
\thispagestyle{empty}
\pagestyle{empty}

\begin{abstract}
    
Brain computer interface (BCI) is the only way for some special patients to communicate with the outside world and provide a direct control channel between brain and the external devices. As a non-invasive interface, the scalp electroencephalography (EEG) has a significant potential to be a major input signal for future BCI systems. Traditional methods only focus on a particular feature in the EEG signal, which limits the practical applications of EEG-based BCI.
In this paper, we propose a algorithm for EEG classification with the ability to fuse multiple features. First, use the common spatial pattern (CSP) as the spatial feature and use wavelet coefficient as the spectral feature. Second, fuse these features with a fusion algorithm in orchestrate way to improve the accuracy of classification. 
Our algorithms are applied to the dataset IVa from BCI complete \uppercase\expandafter{\romannumeral3}. By analyzing the experimental results, it is possible to conclude that we can speculate that our algorithm perform better than traditional methods.

\end{abstract}

\section{INTRODUCTION}

In recent years, BCI has been one of the most interesting biomedical engineering research fields. For patients suffering from stroke, it is very meaningful to provide a communication method to deliver brain messages and commands to the external world except from the normal nerve-muscle output pathway \cite{wolpaw2002brain}. 
Due to the natural and non-intrusive characteristics, most BCI systems select the EEG signal as input.  
Event-related desyn-chronization/synchronization (ERD/ERS) is the Rhythmic activity of EEG. The changes of ERD/ERS induced by Motor Imagery (MI) has been widely used in many BCI systems.
The very aim of EEG signal processing in BCI system is translating raw EEG signals into the class of these signals. In addition, EEG-based BCI can be considered as a complex pattern recognition system which includes several critical steps, signal acquisition, preprocess, feature extraction, classification, control external devices and feedback.
If the EEG signal is decoded correctly, it can be translated into commands to control external equipments such as rehabilitation devices \cite{lew2006non} or other devices  \cite{doud2011continuous}.

In order to improve the accuracy of EEG classification, a lot of related work  has been carried out. The performance of a pattern recognition like system depends on both the features selected and the classification algorithms employed. 
A great variety of features have been proposed to used in BCI such as Band Powers (BP) \cite{kaiser2011first}, Power Spectral Density (PSD) values \cite{waldert2009review}, iterative two-dimensional nearest feature line (INFL) and so on. In recent years, the common spatial pattern (CSP) \cite{ramoser2000optimal} has been proved to be an expressive feature of EEG signals. A lot of related work has been proposed such as CSSP \cite{lemm2005spatio}, CSSSP \cite{dornhege2006combined}, FBCSP \cite{ang2008filter}, WCSP \cite{mousavi2011wavelet} and SCSSP \cite{aghaei2016separable}. 
Moreover, many machine learning techniques such as naive Bayes, K-NN, SVM, LDA, ANN, and Deep Learning \cite{jirayucharoensak2014eeg} have been applied as BCI classifiers.
Although the number of research studies on EEG-based BCI has been great increasing in past decades, it is very difficult to decode the useful information from EEG signal in a reliable and efficient way. 
Most of traditional methods focus on a particular feature only which limited the efficiency of BCI. 

The rest of this paper is organized as follows. Our algorithm will be introduced in Section 2. In Section 3, experimental results and performance analysis are provided. Finally, the conclusions will be given in Section 4.

\section{Method}

\begin{figure}[thpb]
    \centering
    \includegraphics[width=3.25in]{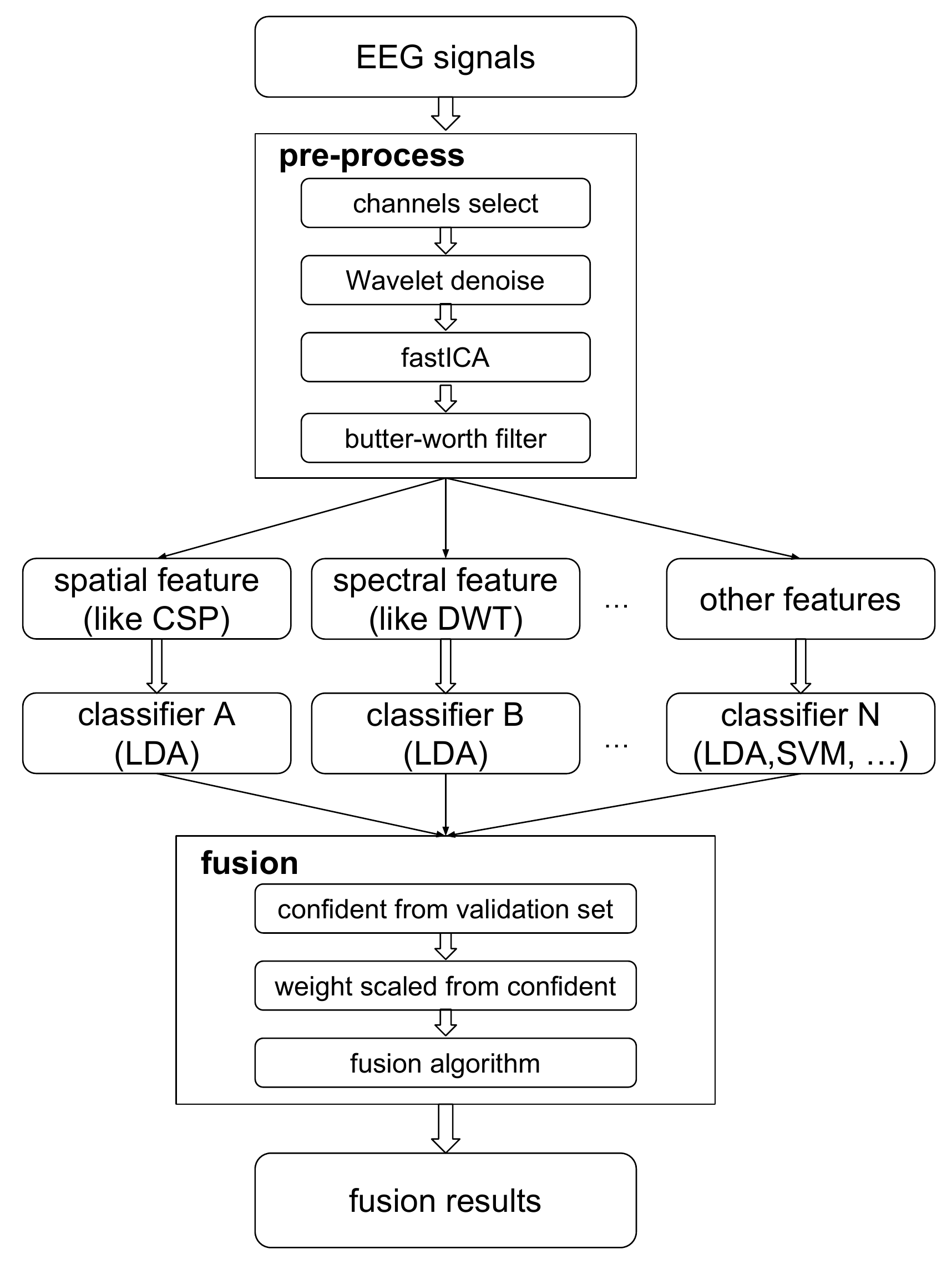}
    \caption{Skeleton of our algorithm}
    \label{our algorithm profile}
\end{figure} 

In this paper, we divide the algorithm into three main stages. The skeleton of our algorithm (stage1-stage3) is schematically illustrated in Fig. \ref{our algorithm profile}. Due to the mass noise in the EEG signal, the preprocess stage is very significant for the following stages. The second stage called feature extraction and classification. In this stage, we extract the spatial features and spectral features, and then use a appropriate classifier called linear discriminant analysis (LDA) to decode the information from these signals. The last stage is fusion processing that obtain the final classification result from the above stages.

\subsection{Preprocess} 
The signals from the different channels correspond to the different functional regions of the human brain. In order to have a positive effect in the subsequent classification algorithm, select some special channels in carefully to obtain a better signal-noise ratio \cite{xu2013channel}. 
In addition, the begin and end portions of the signal segment have been cut off to further improve the signal-to-noise ratio.

Wavelets denoising has been proven to be a powerful tool for removing noise from digital signals. We utilize the wavelet transform with symlets wavelet for improving the signal-noise ratio, and use variant of the first option as threshold selection rule compare to the universal threshold $\sigma \sqrt{2ln(\cdot )}$ that uses in many other BCI systems. The experiments results show that heursure threshold selection rule can yield better results following the steps discussed in the following subsections.

Independent component analysis (ICA) is an effective algorithm for finding fundamental factors or components from a composite signal. FastICA \cite{delorme2004eeglab} is a widely used ICA-like algorithm, which can be more efficient than traditional conventional gradient descent methods. Therefore, we apply the fastICA algorithm to remove noises from other parts of the body and reduce dimension of these signals. 
At the end of the preprocessing, the butter-worth filter has been used as a bandpass filter to remove high-order noise in the signal.

\subsection{Feature extraction}

To improve the accuracy of classification, we use the common spatial pattern of spatial information \cite{ramoser2000optimal} and statistical values of wavelet transform coefficients as the spectral information.

For extracting spectral information, discrete wavelet transform (DWT) has been applied to each fragment in each channel \cite{robinson2013eeg}. In our work, the symlet wavelets have been used as the mother wavelet to generate the wavelet coefficients.
We can simply use the mean and variance of each fragment as an approximation of the spectral feature to obtain the statistical features of spectral information.

\subsection{Classification}
The classification stage in the BCI system is intended to label the feature vector. We use LDA to classify spatial information and spectral information \cite{lotte2007review} \cite{liu2015uncorrelated}. The purpose of LDA is to separate the training feature vectors representing different classes using the hyperplane to obtain the smallest intraclass dispersion and the maximum interclass dispersion.
\begin{eqnarray}
\begin{aligned}
\hat{\varphi} & = \mathrm{argmax} ~J_{fisher}(\varphi) \\
& = \mathrm{argmax} ~\frac{{\varphi}^{T} S_b \varphi}{{\varphi}^{T} S_w \varphi} 
\end{aligned}
\end{eqnarray}
where $S_b$ is the discrete degree between classes, $S_w$ is the discrete degree within the class and $\hat{\varphi}$ is the optimal projection matrix. Naturally, LDA can achieve good result on unseen data, and also good results can be obtained in practical applications.

\subsection{Fusion}

To overcome the limitations and shortcomings of conventional EEG classification algorithms, we use a fusion algorithm based on Bayesian-decision \cite{daunizeau2007symmetrical} to fuse the results of multiple classifiers in different aspects. In a normalized BCI system, there is no favoritism of any class, so we choose the 0-1 lost function to minimize the total error risk. Then Bayesian rule with the smallest error rate using the 0-1 lost function can be formulated as:
\begin{eqnarray}
\begin{aligned}
\hat{j} & = \underset{j \in C} {\mathrm{argmax}} ~P(\omega_i|x) \\  
& = \underset{j \in C} {\mathrm{argmax}} ~P(x|\omega_j)P(\omega_j)
\end{aligned}
\end{eqnarray}
where $\hat{j}$ is the class with the largest possible and $P(\omega_i|x)$ is the probability of classified as $i$ which determined by one classifier. Assuming that there are $R$ EEG classifiers which independent of each other and suppose there is no favoritism of any class, the fusion rule will be rewritten as follow:
\begin{eqnarray}
\begin{aligned}
\label{multiply_classifiers_reduct}
\hat{j} & = \underset{j \in C} {\mathrm{argmax}} ~P(\omega_j|x_1,x_2,...,x_R)\\ 
& =  \underset{j \in C} {\mathrm{argmax}} ~{P(x_1,x_2,...,x_R|\omega_j)P(\omega_j)} \\
& =  \underset{j \in C} {\mathrm{argmax}} ~{P(\omega_j) \prod_{i=1}^{R}{P(x_i|\omega_j)}} \\
& =  \underset{j \in C} {\mathrm{argmax}} ~{P(\omega_j) \prod_{i=1}^{R}{\frac{P(\omega_j|x_i) * p(x_i)}{P(\omega_j)}}} \\
& =  \underset{j \in C} {\mathrm{argmax}} ~{P^{-(R-1)}(\omega_j) \prod_{i=1}^{R}{P(\omega_j|x_i)}}
\end{aligned}
\end{eqnarray}
Suppose that there is only tiny difference between $P(\omega_k)$ and $P(\omega_k|x_j)$, which means:
\begin{equation}
\label{tiny_different_between}
P(\omega_k|x_j) = P(\omega_k)(1+\delta_{kj}), ~~\delta_{kj} << 1
\end{equation}
Ignore items with order greater than 2, then Eq. \eqref{multiply_classifiers_reduct} can be revised as:
\begin{eqnarray}
\begin{aligned}
\hat{j} & =  \underset{j \in C} {\mathrm{argmax}} ~{P^{-(R-1)}(\omega_j) \prod_{i=1}^{R}{P(\omega_j|x_i)}}\\ 
& =  \underset{j \in C} {\mathrm{argmax}} ~{P(\omega_j) \prod_{i=1}^{R}{ (1+\delta_{kj}) }} \\
& \approx  \underset{j \in C} {\mathrm{argmax}} ~{P(\omega_j) (1+\sum_{k=1}^{R}{\delta_{kj}}) } 
\end{aligned}
\end{eqnarray}
We use the accuracy obtained on the validation set as the confidence level of the classifier and set the weight of the classifier to a transformation of the confidence level by a $tan$ like function. Then introduce a scaling factor $s$ to enhance this validity, and the weight $W_i$ will be determined to:
\begin{eqnarray}
\begin{aligned}
W_i^{'} &= ( sign(C_i-\frac{1}{2}) ) \frac{(\frac{1}{2} - \left | C_i-\frac{1}{2} \right |) (s-1)} {s} \\
W_i &= \tan{ 
    \left( \frac{\pi}{2} 
        \left( 
            C_i + W_i^{'}
        \right) 
    \right)
}
\end{aligned}
\end{eqnarray}
where $s \geq 1$ is the scaling factor and $C_i$ is the confidence of the classifier calculated from the validation set.
As we have derived in equation \ref{tiny_different_between}, assume that each class has the same prior probability in our work. The following equation is then used to determine the final fusion result. 
\begin{eqnarray}
\begin{aligned}
\hat{j} & =  \underset{j \in C} {\mathrm{argmax}} ~P(\omega_j|x_1,x_2,...,x_R)\\ 
& = \underset{j \in C} {\mathrm{argmax}} ~P(\omega_j) \sum_{i=1}^{R}{ W_i P(\omega_j|x_i) } 
\end{aligned}
\end{eqnarray}
where $W_i$ is the weight of the classifier estimated in the above steps, and $R$ is the number of classifiers.

\section{EXPERIMENT}

We performed our experiments at the data set IVa that provided by BCI competitions \uppercase\expandafter{\romannumeral3}. 
This data set contains 118 channels of EEG sampled signals which divided into five subjects (with named aa, al, av, aw, ay). Each subject contains 280 segments of data.  The label of each piece of data shows that the subject is doing motor image by performing one of the following three movements: (L) left hand, (R) right hand, (F) right foot.

\begin{figure} 
    \centering 
    \subfigure[before]{ 
        \label{fig:experiment_preprocess:before} 
        \includegraphics[width=2.4in]{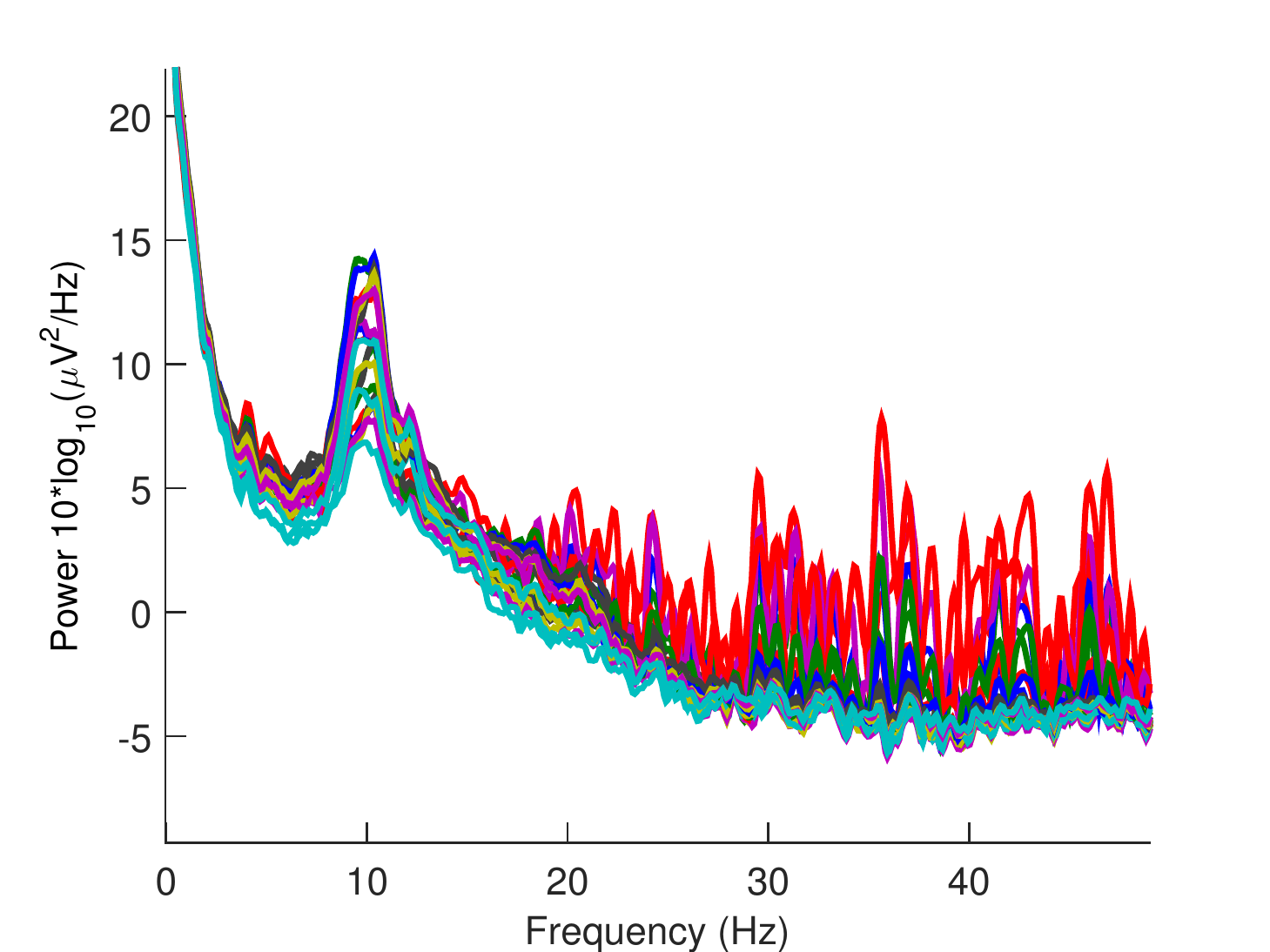}} 
    \hspace{1in} 
    \subfigure[after]{ 
        \label{fig:experiment_preprocess:after} 
        \includegraphics[width=2.4in]{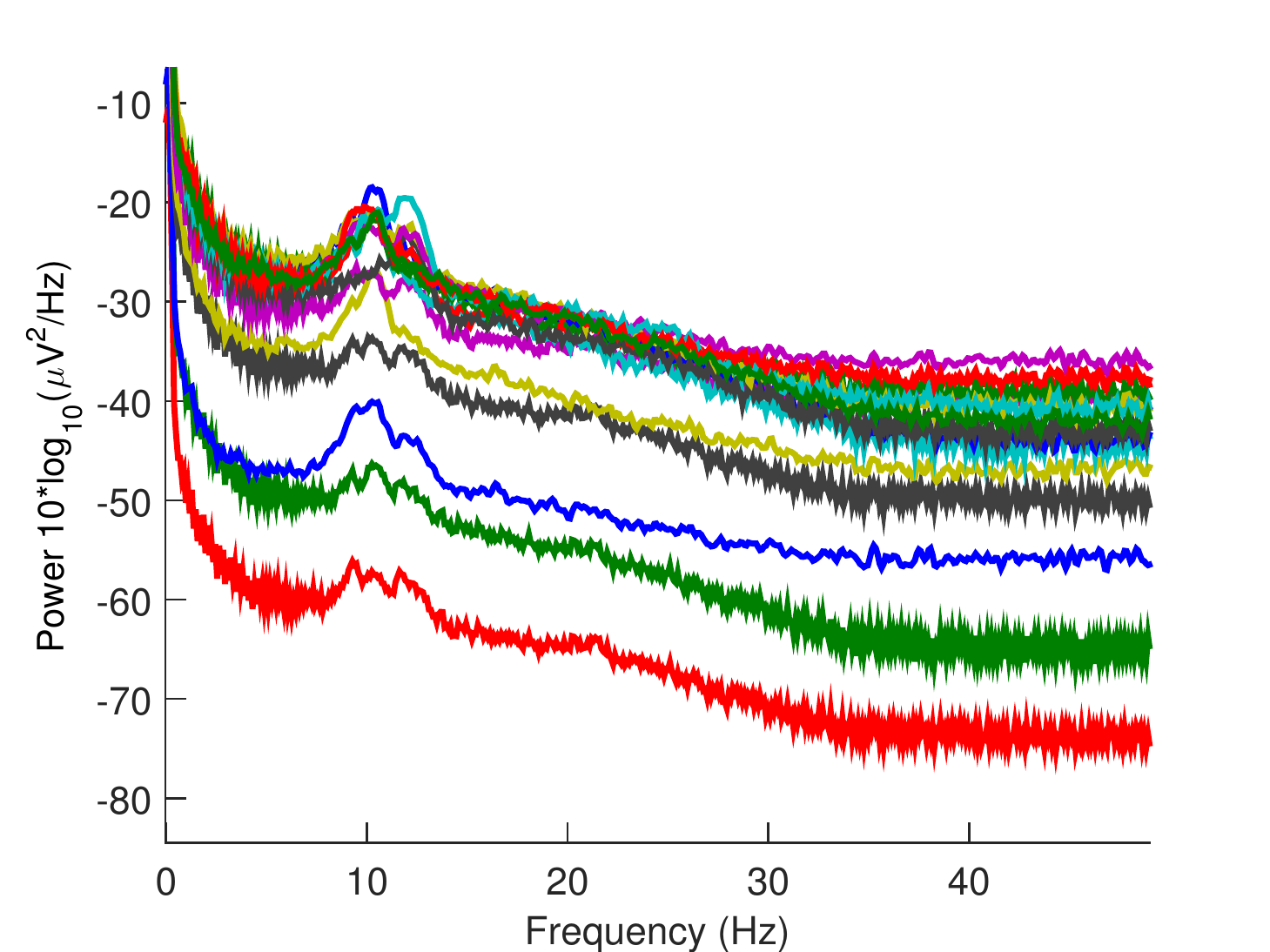}} 
    \caption{Frequency power variation in preprocess} 
    \label{fig:experiment_preprocess} 
\end{figure}

First of all, we choose the 21 channels (CP6, CP4, CP2, C6, C4,
C2, FC6, FC4, FC2, CPZ, CZ, FCZ, CP1, CP3, CP5, C1, C3, C5, FC1, FC3, FC5) from the data set. Second, we carry out the above-mentioned preprocess operation. The influence of preprocessing step on frequency spectrum is shown in Fig. \ref{fig:experiment_preprocess}. This figure clearly shows that our preprocessing operation is useful for feature extraction and classification. Moreover, it can be seen that the useful part of signal after preprocess is concentrated at 8-30 Hz.

The accuracy of classification algorithm can be calculated simply by:

\begin{equation}
accuracy = \frac{N_{right}}{N_{testset}} 
\end{equation}

The experimental results show that our work achieved better result than other algorithms such as PSD, INFL, CSP, CSSP. Experimental results are shown in Table \ref{table: experiment result of different algorithms compare}, and these results demonstrate that our algorithm can achieve a better accuracy in most cases and obtain the best mean value and acceptable variance value.

\begin{table}[h!]
\caption{The experimental results (\%) obtained from each subject using different algorithms}
\label{table: experiment result of different algorithms compare}
\begin{center}
	\begin{tabular}{||c | c cl c c||} 
		\hline
		~     & PSD   & INFL  & CSP   & CSSP & Our Work  \\ [0.5ex] 
		\hline\hline
		   aa & 64.29 & 75.89 & 71.43 & 77.68 & \textbf{85.27}  \\ 
		\hline
		   al & 82.14 & 83.93 & 94.64 & 96.43 & \textbf{100}  \\
		\hline
		   av & 55.10 & 58.16 & 61.22 & 63.27 & \textbf{69.89} \\
		\hline
		   aw & 65.18 & 82.14 & 89.28 & \textbf{90.63} & 85.71 \\
		\hline
           ay & 75.00 & 84.92 & 73.02 & 79.37 & \textbf{82.34} \\
        \hline
         mean & 68.34 & 77.00 & 77.92 & 81.48 & \textbf{84.64}   \\
         \hline
         var  & \textbf{87.3}& 98.67& 150.73& 131.54& 92.06\\
 [1ex] 
		\hline
	\end{tabular}
\end{center}
\end{table}

\begin{figure}
    \centering
    \subfigure[segment size of signal]{
        \includegraphics[width=1.56in]{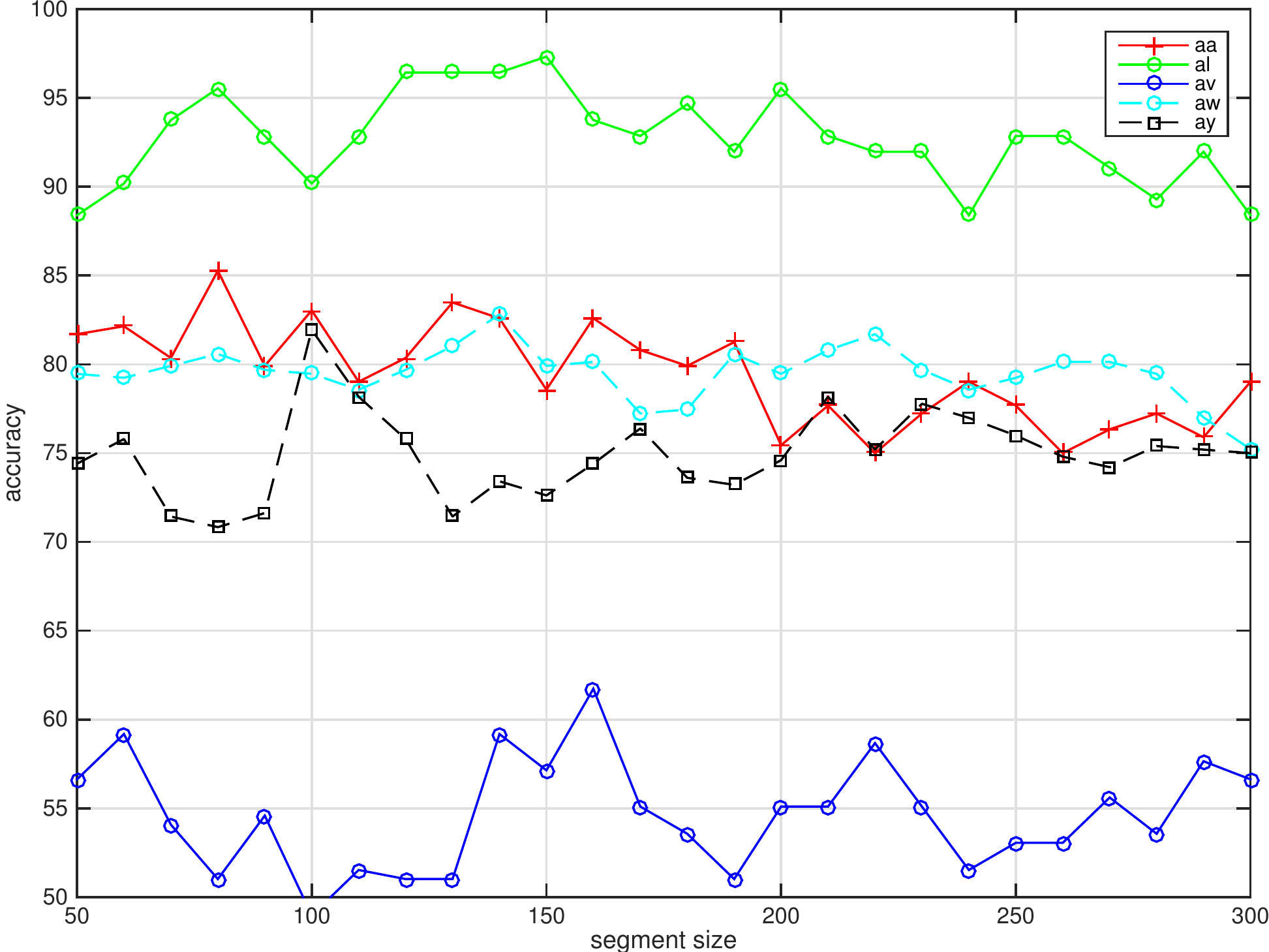}
        \label{figure:parameter_iterator:a}
    }
    \vspace{0in} \hspace{0in}
    \subfigure[dimension of ICA]{
        \includegraphics[width=1.56in]{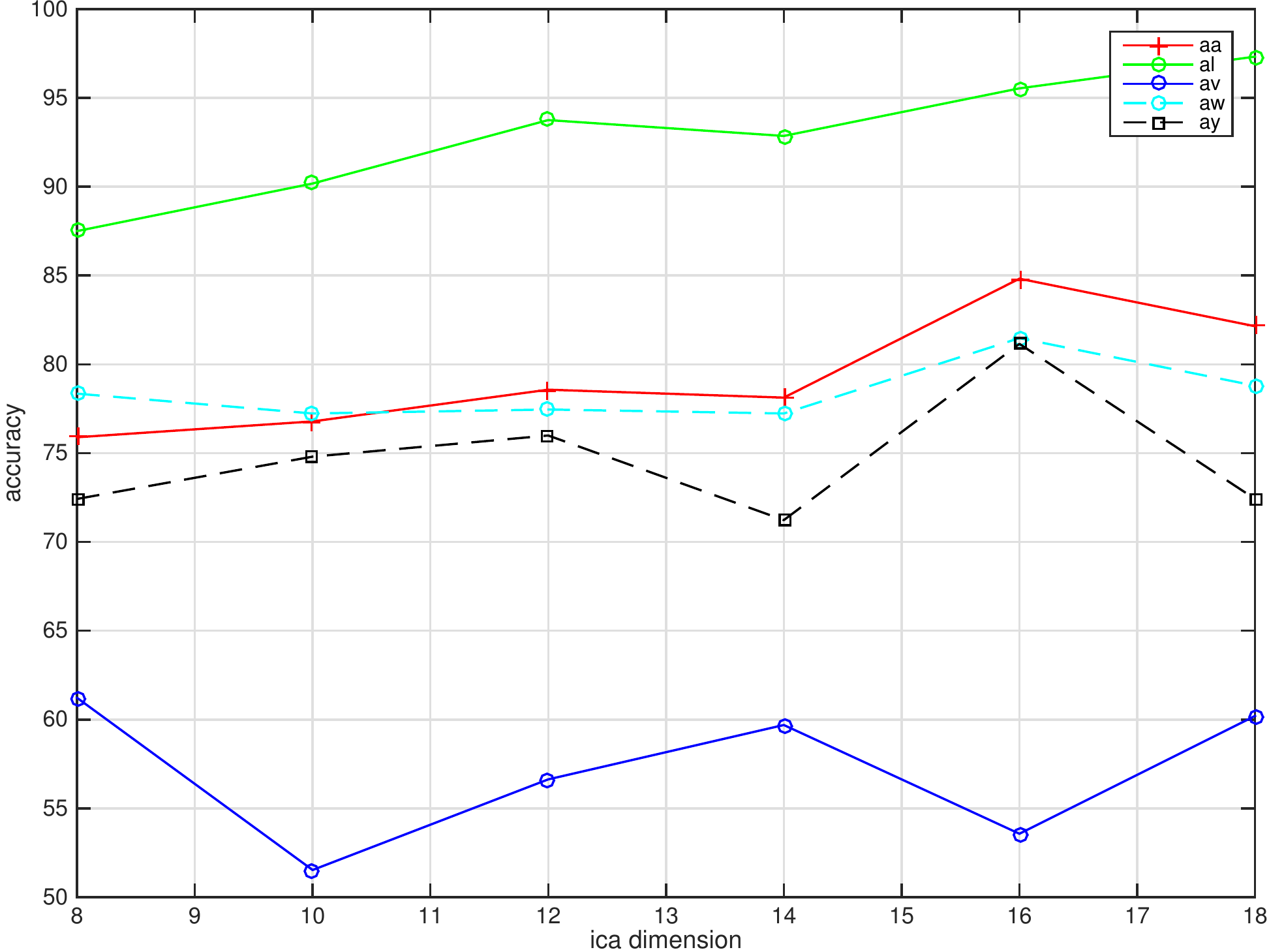}
        \label{figure:parameter_iterator:b}
    }
    
    \subfigure[size of CSP]{
        \includegraphics[width=1.56in]{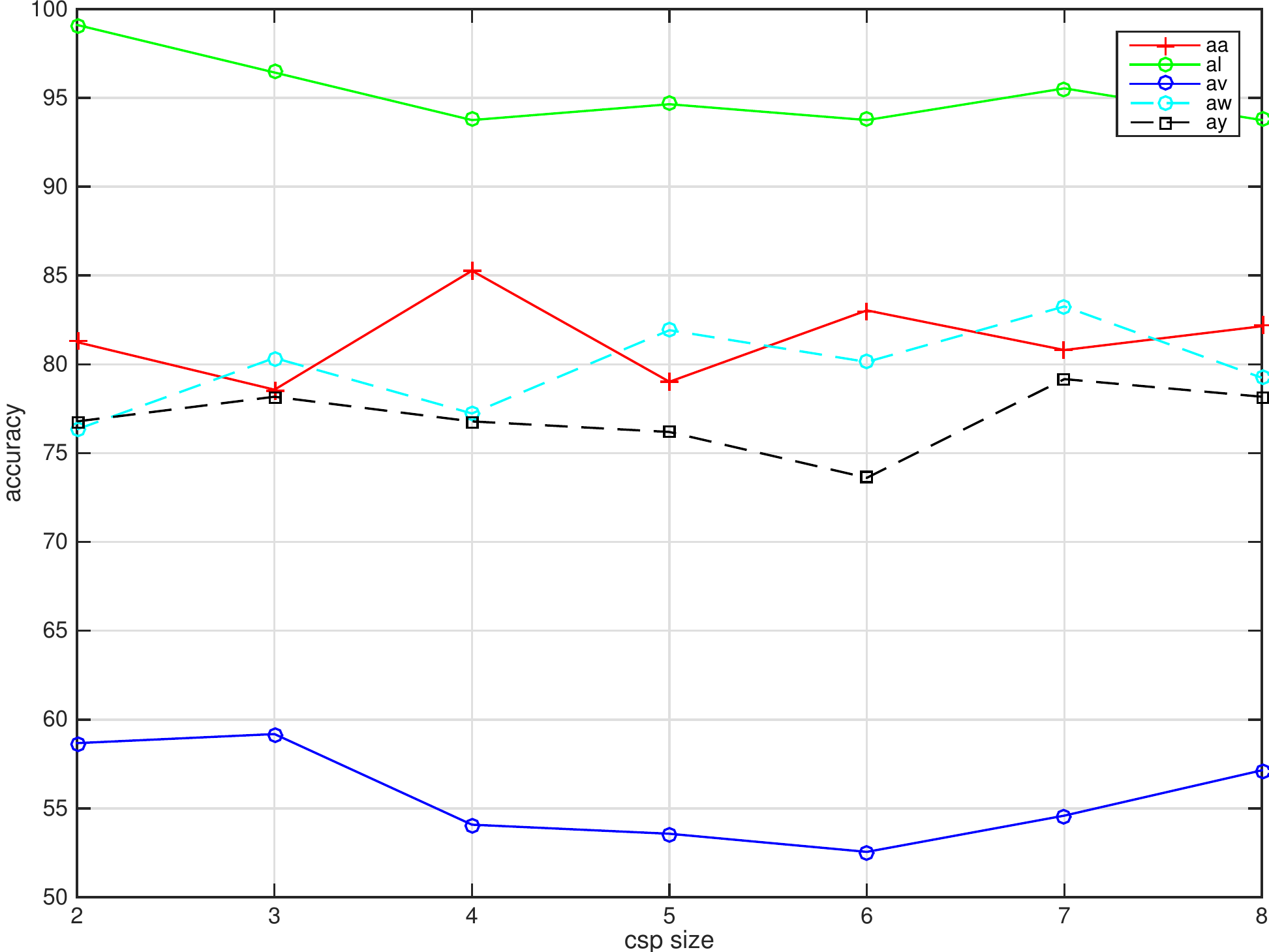}
        \label{figure:parameter_iterator:c}
    }
    \vspace{0in} \hspace{0in}
    \subfigure[order of butter-worth filter]{
        \includegraphics[width=1.56in]{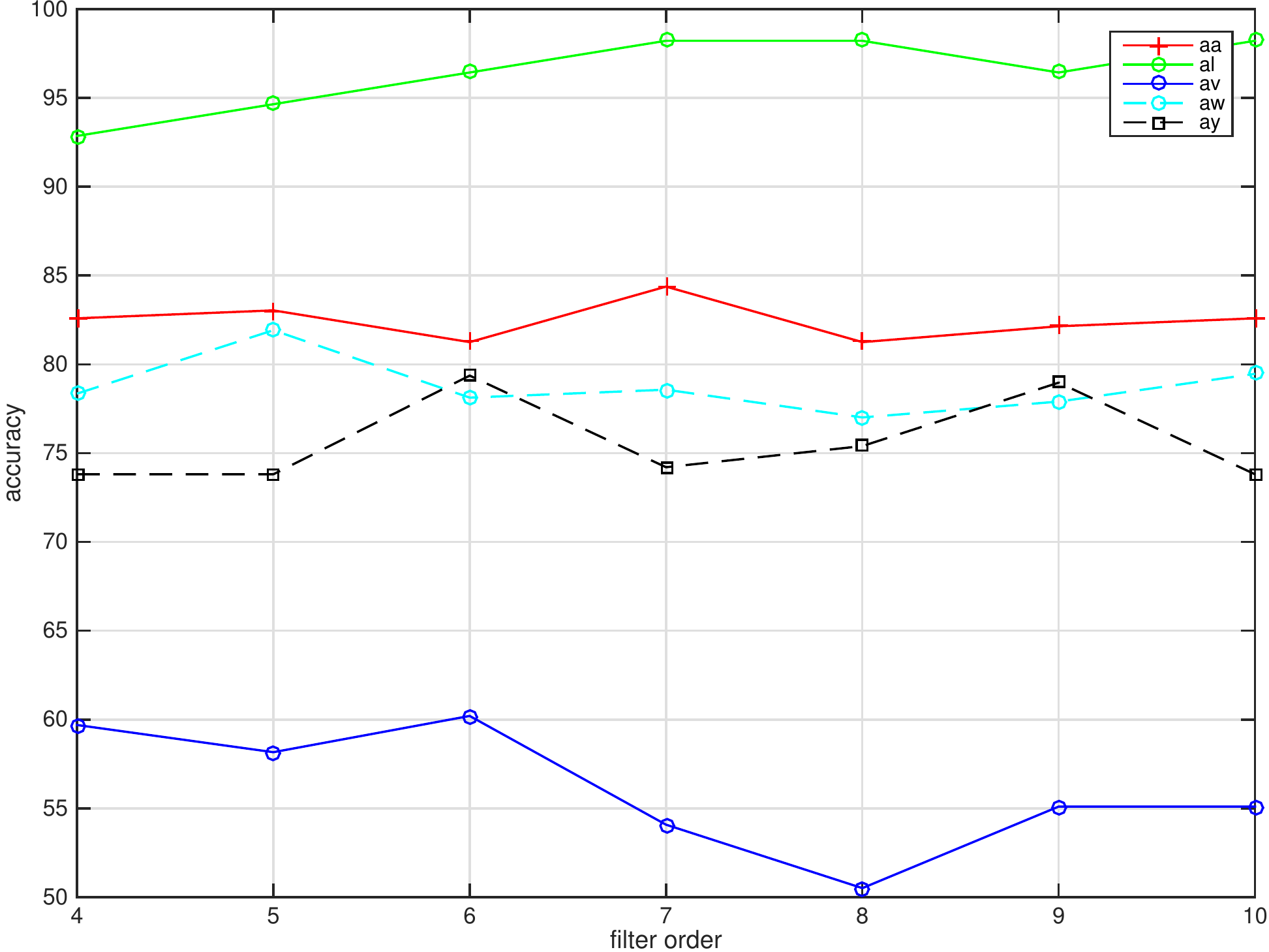}
        \label{figure:parameter_iterator:d}
    }
    \caption{Accuracy curves obtained using different parameter values}
    \label{figure:parameter_iterator}
\end{figure}

In order to optimize the parameters of the algorithm in the experiments, reasonable range of each parameter in our algorithm has been pointed. The classification accuracy results with different parameter values are shown in Fig. \ref{figure:parameter_iterator}, and these figures can indicate the following items:

\begin{itemize}
	\item EEG segment size of the reasonable range is 140-240, which means that we should choose the middle of the motor image signal with the length of 1.4s-2.4s.
	\item The best value for the ICA dimension is 16.
	\item The reasonable size of CSP seems not easy to determined, but the appropriate value in most situations is between 2 and 3. 
	\item The reasonable order of butter-worth filter can be determined as 5-6.
\end{itemize}

\begin{table}[h!]
    \caption{Comparison of experiment results (\%) between our fusion classifier and basic spatial classifier}
    \label{table: compare use fusion}
    \begin{center}
        \begin{tabular}{||c | c c||} 
            \hline
            ~  & Basic spatial classifier & Our fusion classifier      \\ [0.5ex] 
            \hline\hline
            aa & 83.48        & \textbf{85.27}         \\ 
            \hline
            al & 97.32        & \textbf{100}           \\
            \hline
            av & 60.20        & \textbf{69.89}         \\
            \hline
            aw & 80.13        & \textbf{85.71}         \\
            \hline
            ay & 78.17        & \textbf{82.34}         \\
            \hline
            mean & 79.86        & \textbf{84.64}         \\
            \hline
            var & 141.48       & \textbf{92.06}        \\ [1ex] 
            \hline
        \end{tabular}
    \end{center}
\end{table}

For the fusion stage, we compare the accuracy of our fusion classifier with the basic CSP classifier, and the results are shown in Table \ref{table: compare use fusion}. Then, try to optimize the scale factor in the fusion algorithm and find that the reasonable range for this parameter can be determined as 1.4-2.4. These results show that our fusion algorithm plays a positive role in classification.

\section{CONCLUSIONS}
Realizing the information exchange between the brain and the computer based on EEG signal is still a challenge in signal processing and biomedical engineering. 
In this paper, our algorithm has been proposed to overcome the limitations and shortcomings of conventional EEG classification algorithms with multiple features. 
The signal is preprocessed to improve quality, represented by multiple features in different aspects, and fed to LDA classifiers. 
The results of these classifiers is to determine the final classification result of the signal by our fusion algorithm.
Experimental results have shown that the proposed approach in this paper is an effective way to classify EEG signal in a BCI system.

\section{ACKNOWLEDGE}

This work was supported by the National Natural Fund: 91420302 and 91520201. Thanks to the organizers of the BCI competition \uppercase\expandafter{\romannumeral3} to publish their datasets.

\addtolength{\textheight}{-12cm}   

\bibliographystyle{unsrt}
\bibliography{myreference}

\end{document}